\documentclass[aps,pra,11pt,superscriptaddress,nofootinbib]{revtex4-2}
\newcommand{\figurewidth}{0.6\linewidth}
\usepackage{newtxtext,newtxmath}

\usepackage{slashed}
\usepackage[section]{placeins}
\usepackage{graphicx,enumitem}
\usepackage[normalem]{ulem}
\usepackage[dvipsnames]{xcolor}
\usepackage[colorlinks,allcolors=Blue,linktocpage]{hyperref}
\usepackage{CJK,calrsfs}
\newcommand{\SO}{\mathrm{SO}}
\newcommand{\SU}{\mathrm{SU}}
\newcommand{\Sp}{\mathrm{Sp}}
\newcommand{\rO}{\mathrm{O}}
\newcommand{\rU}{\mathrm{U}}
\newcommand{\rd}{\mathrm{d}}
\newcommand{\br}{\mathbf{r}}
\newcommand{\bbI}{\mathbb{I}}
\newcommand{\bbZ}{\mathbb{Z}}
\newcommand{\cL}{\mathcal{L}}
\newcommand{\hc}{\text{h.c.}}
\newcommand{\CS}{\text{CS}}
\DeclareMathOperator{\tr}{tr}
\DeclareMathOperator{\sgn}{sgn}

\begin{document}

\begin{CJK*}{UTF8}{bkai}

\title{Chern-Simons-matter conformal field theory on fuzzy sphere: \\Confinement transition of Kalmeyer-Laughlin chiral spin liquid}

\author{Zheng Zhou (周正)}
\affiliation{Perimeter Institute for Theoretical Physics, Waterloo, Ontario N2L 2Y5, Canada}
\affiliation{Department of Physics and Astronomy, University of Waterloo, Waterloo, Ontario N2L 3G1, Canada}
\author{Chong Wang}
\affiliation{Perimeter Institute for Theoretical Physics, Waterloo, Ontario N2L 2Y5, Canada}
\author{Yin-Chen He}
\affiliation{C. N. Yang Institute for Theoretical Physics, Stony Brook University, Stony Brook, NY 11794-3840}

\begin{abstract}
    Gauge theories compose a large class of interacting conformal field theories in 3d, among which an outstanding category is critical Chern-Simons-matter theories. In this paper, we focus on one of the simplest instances: one complex critical scalar coupled to $\rU(1)_2$ Chern-Simons gauge field. It is theoretically interesting as it is conjectured to exhibit dualities between four simple Lagrangian descriptions, but also practically important as it describes the transition between Kalmeyer-Laughlin chiral spin liquid (or $\nu=1/2$ bosonic Laughlin state) and trivially gapped phase. Using the fuzzy sphere regularisation, we realise this theory as a transition on the spherical lowest Landau level between a $\nu_f=2$ fermionic integer quantum Hall state and a $\nu_b=1/2$ bosonic fractional quantum Hall state. We show that this transition is continuous and has emergent conformal symmetry. By studying the operator spectrum, we show that there exists only one relevant singlet with scaling dimension $\Delta_S=1.52(18)$. We also discuss other higher operators and the consequences of our results.
\end{abstract}

\date{19 August, 2025}

\maketitle

\end{CJK*}

\textit{Introduction} ---
Conformal field theory (CFT) is a central topic of modern condensed matter~\cite{Polyakov1970Conformal,Cardy1996Scaling} and high energy physics~\cite{Maldacena1998AdSCFT}. While many 2d CFTs are exactly solvable~\cite{Belavin1984BPZ,DiFrancesco1997CFT}, the solution to 3d CFTs remains an open and important problem. In 3d, coupling matter to gauge field is one of few known ways to construct interacting critical theories. Critical gauge theories also describe many critical phases and phase transitions in the infrared. One outstanding catagory is the Chern-Simons (CS) matter theories, which have critical fermions/scalars coupled to CS gauge field. The CS-matter theories describe transitions between, or out of, topological orders~\cite{Wen1993transitions,Chen1993Mott,Barkeshli2012FQH,Lee2018QEDCS,Zou2018QCD,Samajdar2019Duality,Ma2020QCD,Song2022CSL,Song2023Moire,Wu2023FQHSPT,Goldman2019NAQH,Shankar2022CSL}. The recent developments in Moir\'e materials provide an exciting opportunity to study these transitions experimentally~\cite{Cai2023Moire}. Theoretically, they also enjoy interesting properties such as field theory dualities~\cite{Hsin2016LevelRank,Seiberg2016Duality,Benini2017Duality,Son2015Duality,Metlitski2015Duality,Wang2015Duality,Karch2016Duality,Aharony2016Duality}. Despite these theoretical and experimental significance, it is difficult to study their nature, such as critical exponents, since quantum Monte Carlo suffers from sign problem in the absence of parity symmetry. 

Among these CS-matter theories, one of the simplest is the confinement transition of the Kalmeyer-Laughlin chiral spin liquid (KL-CSL). This theory is conjectured to have $\SO(3)$ global symmetry in the IR as well as four Lagrangian descriptions that are dual to each other: $\SU(2)_1$ with a scalar, $\rU(1)_2$ with a scalar, $\SU(2)_{-1/2}$ with a fermion, and $\rU(1)_{-3/2}$ with a fermion~\cite{Hsin2016LevelRank,Seiberg2016Duality,Benini2017Duality}. It is commonly believed to describe a continuous phase transition between a trivially gapped phase and a semion topological order. In various contexts, this putative transition can happen between a KL-CSL~\cite{Kalmeyer1987CSL,Kalmeyer1989CSL,Wen1989CSL,He2013CSL,Bauer2014CSL,Gong2014,Szasz2018CSL,Wietek2016CSL,Kuhlenkamp2022CSL,Divic2024CSL} and a trivially gapped state, or between a bosonic fractional quantum Hall state at filling $\nu=1/2$~\cite{Laughlin1983FQHE} and an integer quantum Hall state. Upon doping at the critical point, it is recently argued that topological superconductivity emerges, where the lowest charge excitation is dominated by bosonic charge-2 modes (\textit{i.e.}, fermion pairing)~\cite{Divic2025AnyonSC}. Similar anyon superconductivity has also been studied in the vicinity of other transitions~\cite{Minho2025AnyonSC,Shi2024AnyonSC,Pichler2025AnyonSC,Zhang2025AnyonSC,Shi2025AnyonSC,Nosov2025AnyonSC,Wang2025SC}. Despite the broad interest in this transition, its nature is poorly understood.

The fuzzy sphere regularisation~\cite{Zhu2022} has emerged as a new powerful method to study 3d CFTs~\cite{Zhu2022,Hu2023Mar,Han2023Jun,Zhou2023,Hu2023Aug,Hofmann2023,Han2023Dec,Zhou2024Jan,Hu2024,Cuomo2024,Zhou2024Jul,Dedushenko2024,Fardelli2024,Fan2024,Zhou2024Oct,Voinea2024,Yang2025Jan,Han2025,Laeuchli2025,Fan2025,ArguelloCruz2025,EliasMiro2025,He2025,Taylor2025,Yang2025Jul}. This approach allows manifest exact rotation symmetry and direct extraction of conformal data through state-operator correspondence. In a previous work~\cite{Zhou2024Oct}, we have discovered a series of new $\Sp(N)$-symmetric CFTs. By matching through Wess-Zumino-Witten (WZW) models~\cite{Komargodski2017QCD,Lee2014WZW}, we have shown that their candidates are CS-matter theories. However, it is difficult to identify its phase diagram and corresponding field theory description. On the other hand, while most of the existing models on the fuzzy sphere are at integer filling with quantum Hall ferromagnet part of the phase diagram, Voinea \textit{et al.}~\cite{Voinea2024} have recently shown that the fuzzy sphere regularisation can also be applied to charge-gapped fractional quantum Hall phases. This advance sheds light on new possibility of realising the CS-matter theories as the phase transitions between integer or fractional quantum Hall states. This setup provides better guidance to target the theory of interest.

In this paper, we study the confinement transition of KL-CSL. Our setup involves two flavours of charge-1 fermions in the fundamental representation of $\SU(2)$ flavour symmetry and one flavour of charge-2 bosons. The conversion between two fermions and one boson is allowed. By tuning a relative chemical potential, a phase transition between a fermionic integer quantum Hall phase with filling $\nu_f=2$ and a bosonic fractional quantum Hall phase with filling $\nu_b=1/2$ is reached, as originally studied on a disk geometry~\cite{Yang2008Feshbach,Liou2018Feshbach}. We show that this transition can be captured by several Chern-Simons matter theories that are dual to each other. Thanks to the spherical geometry, we obtain clear numerical evidence that the phase transition is continuous and has emergent conformal symmetry. By studying the operator spectrum, we show that there exists only one relevant primary operator (except symmetry current) that is a symmetry singlet and has scaling dimension $\Delta_S=1.52(18)$. We also discuss other higher operators. This finding lays foundation for researches related to this phase transition like anyon superconductivity and other CSL-related transitions. Our work opens up a new route for studying the nature and the conformal data of critical gauge theories.

\begin{figure}[t]
    \centering
    \includegraphics[width=\figurewidth]{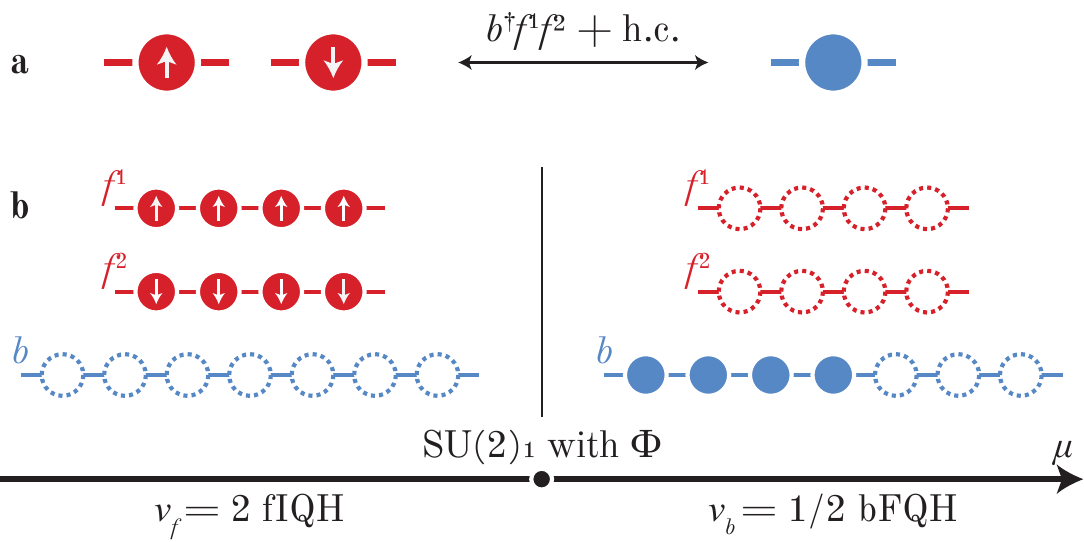}
    \includegraphics[width=\figurewidth]{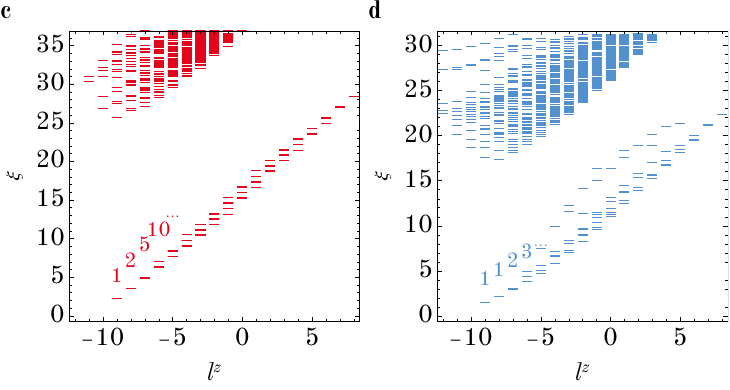}
    \caption{(a) A sketch of the conversion between two fermions and a boson. (b) A sketch of the setup on the fuzzy sphere and the phase diagram. (c,d) The entanglement spectrum with real-space cut at the equator in (c) the fIQH phase at $\mu=-2,N_{mf}=6$ and (d) the bFQH phase at $\mu=2,N_{mf}=6$. The degeneracy of the edge mode is marked.}
    \label{fig:1}
\end{figure}
 
\textit{Model} ---
In the setup of the fuzzy sphere~\cite{Madore1991Fuzzy}, we consider a sphere with a $4\pi q$-monopole at its centre, and put two flavours of charge-1 fermion $f^i(\br)$ ($i=1,2$) in the fundamental representation of $\SU(2)$ flavour symmetry and one flavour of charge-2 boson $b(\br)$ on the sphere. Due to the presence of the monopole, the single-particle eigenstates form highly degenerate quantised Landau levels~\cite{Haldane1983FQHE}. The lowest Landau level (LLL) has angular momentum $q$ and degeneracy $N_{mf}=2q+1$ for each fermion flavour and angular momentum $2q$ and degeneracy $N_{mb}=4q+1$ for the boson. By setting the single particle gap the leading energy scale, we project the system onto the LLL and express the fermion and boson operators in terms of the creation and annihilation operators on the LLL 
\begin{equation*}
    f^i(\br)=\frac{1}{R}\sum_{m=-q}^qY_{qm}^{(q)}f^i_m,\quad b(\br)=\frac{1}{R}\sum_{m=-2q}^{2q}Y_{2q,m}^{(2q)}b_m,
\end{equation*}
where the radius is taken as $R=N_{mf}^{1/2}$~\cite{Zhu2022}, and the monopole spherical harmonics can be expressed in terms of the Hopf spinor~\cite{Haldane1983FQHE} $Y_{qm}^{(q)}\propto u^{q+m}v^{q-m}$, and $u=e^{i\phi/2}\sin\theta/2$, $v=e^{-i\phi/2}\cos\theta/2$.

We start with an integer quantum Hall state with fully filled fermion orbitals $\nu_f=2$ and empty boson orbitals. We allow a local conversion between a boson and two fermions $(b^\dagger(\br)f^1(\br)f^2(\br)+\hc)$ that conserves the electric charge (Fig.~\ref{fig:1}a). Physically, this conversion could correspond to the Feshbach resonance where two fermionic atoms combine into a bosonic molecule~\cite{Giorgini2008Feshbach,Yang2008Feshbach,Liou2018Feshbach}. By tuning a relative chemical potential $\mu n_f$, we can convert all the fermions to bosons. In this process, the particle number is halved and the orbital number is doubled, so the boson filling is $\nu_b=\nu_f/4=1/2$~(Fig.~\ref{fig:1}b). This filling allows us to realise a bosonic Laughlin state $\Psi(\{u_i,v_i\})=\prod_{i<j}(u_iv_j-u_jv_i)^2$ by adding a local boson density interaction $n_b(\br)^2$ (\textit{i.e.}, pseudopotential with $l=0$)\footnote{Here the numbers of particles $N_b$ and orbitals $N_{mb}$ are actually related by $2N_b=N_{mb}+1$ due to the Wen-Zee shift~\cite{Wen1992Shift}.}~\cite{Haldane1983FQHE,Laughlin1983FQHE}. This setup was previously studied on the disc geometry by Liou \textit{et al.}~\cite{Liou2018Feshbach}. We also allow a boson-fermion density interaction and a fermion-fermion density interaction. Altogether, the Hamiltonian reads
\begin{equation}
    H=\int\rd^2\br\left[n_e^2-\tfrac{1}{2}(b^\dagger f^1f^2+\hc)+\mu n_f\right],
    \label{eq:hmt}
\end{equation}
where $n_e=n_f+2n_b$ is the electric charge density. A more general form of interaction includes $U_fn_f^2+4U_bn_b^2+4U_{bf}n_bn_f$. The choice $U_f=U_b=U_{bf}=1$ corresponds to $n_e^2$. In this paper, we mainly focus on this simple choice, but will also study the transition at different $U_{bf}$. This model has a $\SU(2)$ flavour symmetry and an electric charge conservation symmetry $\rU(1)_e$ that decouples at the critical point\footnote{As the $\rU(1)_e$ becomes gapped and decouples at the critical point, the $\bbZ_2$ centre of $\SU(2)$ is also projected (\textit{i.e.}, the pseudoreal representation sectors of $\SU(2)$ are gapped), resulting in a global symmetry $\mathrm{PSU}(2)\equiv\SO(3)$. Also note that the symmetry is $\SO(3)$ but not $\rO(3)$ due to the absence of the improper $\bbZ_2$.}.

Tuning the chemical potential $\mu$ drives a phase transition between the $\nu_f=2$ fermionic integer quantum Hall state at large negative $\mu$ characterised by quantum Hall conductance $\sigma_{xy}=2$, spin Hall conductance $\sigma_{xy}^{\SU(2)}=1$ and thermal Hall conductance $\kappa_{xy}=2$, and a $\nu_b=1/2$ bosonic fractional quantum Hall state at large positive $\mu$ with semion topological order and responses $\sigma_{xy}=2,\sigma_{xy}^{\SU(2)}=0,\kappa_{xy}=1$\footnote{Here $\sigma_{xy}$ and $\sigma_{xy}^{\SU(2)}$ are measured in the unit $e^2/h$, and $\kappa_{xy}$ is measured in the unit of $(\pi/6)(k_B^2T/\hbar)$.}. This transition can be captured by four CS-matter theories that are conjectured to be dual~\cite{Hsin2016LevelRank,Benini2017Duality,Seiberg2016Duality}. Among them two are $\SU(2)$ gauge theories
\begin{align}
    \cL&=|(\partial_\mu-i\beta_\mu-iB_\mu)\Phi|^2+\frac{1}{4\pi}\CS[\beta]-m^2|\Phi|^2-\lambda|\Phi|^4\nonumber\\
    \cL&=\bar{\Psi}(i\slashed{\partial}+\slashed{\alpha}+\slashed{B}-m)\Psi-\frac{1}{8\pi}\CS[\alpha]+\frac{1}{8\pi}\CS[B],
    \label{eq:lag_su2}
\end{align}
where $\Psi$ and $\Phi$ are $\SU(2)$-fundamental Dirac fermion and complex scalar, $\alpha_\mu$ and $\beta_\mu$ are dynamical $\SU(2)$ gauge fields and $B_\mu$ is a $\SU(2)$ probing field\footnote{Here $\alpha,\beta$ couple to the gauge degree of freedom, while the probe field $B$ couples to the flavour degree of freedom.}. We use shorthand notations $a_1\,\rd a_2$ for $\epsilon^{\mu\nu\rho}a_{1,\mu}\partial_\nu a_{2,\rho}$ and $\CS[\alpha]$ for $\epsilon^{\mu\nu\rho}\tr(\alpha_\mu\partial_\nu\alpha_\rho+\frac{2}{3}\alpha_\mu\alpha_\nu\alpha_\rho)$. These two Lagrangians have a manifest $\SO(3)$ symmetry, which is the $\SO(3)$ global symmetry of our microscopic model. Another two conjectural dual theories are $\rU(1)$ gauge theories,

\begin{align}
    \cL&=|(\partial_\mu-ib_\mu+iA_\mu)\phi|^2+\frac{2}{4\pi}b\,\rd b-m^2|\phi|^2-\lambda|\phi|^4\nonumber\\
    \cL&=\bar{\psi}(i\slashed{\partial}+\slashed{a}-m)\psi-\frac{3}{8\pi}a\,\rd a-\frac{2}{2\pi}A\,\rd a-\frac{2}{4\pi}A\,\rd A.
    \label{eq:lag_u1}
\end{align}
where $\psi$ is a Dirac fermion, $\phi$ is a complex scalar, $a_\mu$ and $b_\mu$ are $\rU(1)$ dynamical gauge fields. These two Lagrangians only have a manifest $\rO(2)$ global symmetry. The $\rU(1)$ part is the flux conservation of the gauge field, and $A_\mu$ is the probe field for this symmetry. We remark that this $\rU(1)$ symmetry is not the electric charge conservation symmetry of our microscopic model, but a $\rU(1)$ subgroup of the $\SO(3)$ global symmetry\footnote{When the $\rO(2)$ symmetry is enhanced to $\SO(3)$, the $\rU(1)$ symmetry current with $l=1$ and $\rU(1)$ charge $q=0$ combines with the $2\pi$-monopole with $l=1,q=\pm 1$ to form the $\SO(3)$ symmetry current.}. Here we have omitted the gravitational Chern-Simons terms (\textit{cf.} Appendix~\ref{app:a}). For the bosonic theories, $m^2>0$ gaps out the scalar and $m^2<0$ condenses the scalar and Higgses the dynamical gauge field; for the fermionic theories, mass term that gaps out the fermion adds a CS level $\pm 1/2$. Each of the theories flows to a trivially gapped state with $\SU(2)$ Hall conductance $1$ (in Eq.~\eqref{eq:lag_su2}) or $\rU(1)$ Hall conductance $2$ (in Eq.~\eqref{eq:lag_u1}) at $m>0$ and a semion topological order with no response to the probe field at $m<0$. 

\begin{figure}[t]
    \centering
    \includegraphics[width=\figurewidth]{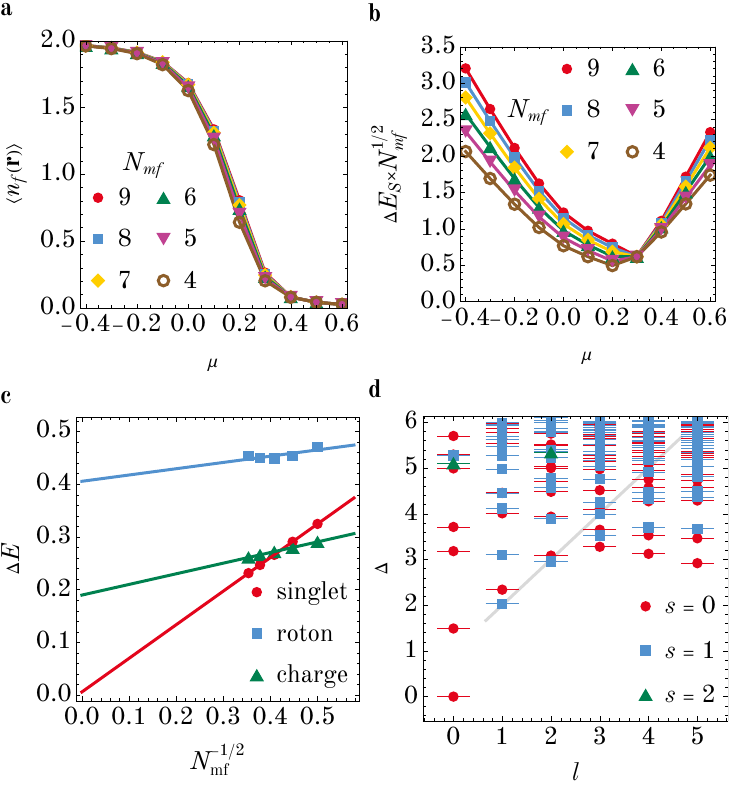}
    \caption{(a) The average fermion density $\langle n_f(\br)\rangle$ as a function of $\mu$ at different sizes, which changes continously. (b) The singlet gap $\Delta E_S$ rescaled with $N_{mf}^{1/2}$ as a function of $\mu$ at different size $N_{mf}$. (c) The finite size scaling of the sclar gap, $l=4$ roton gap and charge gap at the conformal point $\mu=0.312$. (d) The rescaled energy spectrum at the conformal point $\mu=0.312$ and $N_{mf}=9$. The grey line marks the unitarity bound.}
    \label{fig:2}
\end{figure}

\textit{Results} --- 
We numerically calculate this model with exact diagonalisation using our package FuzzifiED~\cite{FuzzifiED} (\textit{cf.} Appendix~\ref{app:b}). The maximal accessible system size is $N_{mf}=9$. At large positive and negative $\mu$ respectively, the $\nu_f=2$ fIQH phase and $\nu_b=1/2$ bFQH phase can be verified by calculating the edge modes~\cite{Wen1990Edge,Wen1995Edge,Li2008Edge,Sterdyniak2011RealEnt}. The real-space entanglement spectra exhibit lowest chiral edge modes whose degeneracies are consistent with one and two chiral free bosons, respectively (Fig.~\ref{fig:1}c,d). In between, the average fermion density $\langle n_f(\br)\rangle$ changes continuously from $0$ to $2$ (Fig.~\ref{fig:2}a), hinting at a continuous phase transition. 

A stronger evidence for the continuous phase transition comes from the closure of energy gap at the critical point. The lowest excited state is a $l=0,s=0$ state\footnote{Here we denote the spin of $\SO(3)$ rotation symmetry by $l$, the spin of $\SO(3)$ flavour symmetry by $s=0$, and $l=0,s=0$ representation by `singlet'.}. We refer to its excited energy as the `singlet gap' $\Delta E_S$. It scales as $1/R\sim N_{mf}^{-1/2}$ at the critical point (Fig.~\ref{fig:2}c), and $\Delta E_SN_{mf}^{1/2}$ increases with system size away from the critical point (Fig.~\ref{fig:2}b). The charge gap, defined as $\frac{1}{2}(\Delta E_{+1}+\Delta E_{-1})$, where $\Delta E_{\pm 1}$ is the excited energy when inserting and removing one fermion, scales to constant value as $N_{mf}\to\infty$ (Fig.~\ref{fig:2}c), confirming that the electric charge conservation symmetry $\rU(1)_e$ is decoupled at the critical point. 

We then take a closer look at the energy spectrum in the charge-neutral sector. The spectrum of a CFT on a sphere enjoys the state-operator correspondence, \textit{i.e.}, every scaling operator has a corresponding eigenstate, and the excited energy $\Delta E=(v/R)\Delta$ is proportional to the scaling dimension $\Delta$ where $v$ is a model-dependent rescaling factor. With proper rescaling, we indeed observe integer-spaced multiplet structure at low energy and $l\leq 2$ (Fig.~\ref{fig:2}d). At larger $l$, there are states that violate the unitarity bound $\Delta\geq l+1$ at small $N_{mf}$. These states are indeed gapped, as confirmed by the finite size scaling taking the lowest $l=4$ excited state as an example (Fig.~\ref{fig:2}c). They are likely to have the similar origin as the magneto-roton modes of the fractional quantum Hall phase in the vicinity~\cite{Girvin1985Roton,Girvin1986Roton}. Since these magneto-roton like states are gapped, eventually at large enough $N_{mf}$ the gapless CFT states will dominate the low energy spectrum at $l>2$. However given that we are only studying small $N_{mf}$, we will focus on the $l\leq 2$ states to analyse the CFT part of the spectrum. 

\begin{figure}[t]
    \centering
    \includegraphics[width=\figurewidth]{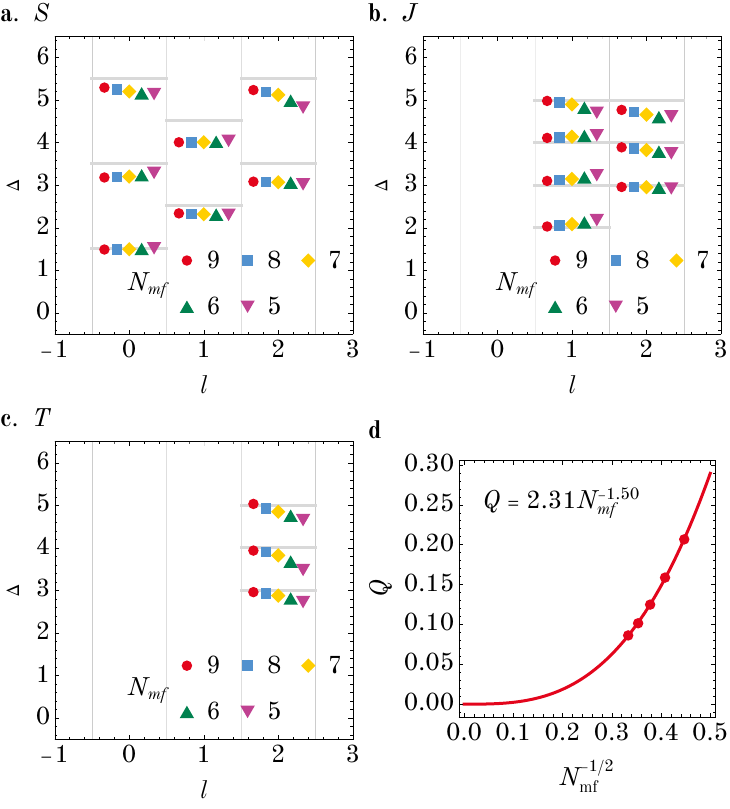}
    \caption{(a--c) The scaling dimensions of the conformal multiplet of (a) the lowest singlet $S$, (b) symmetry current $J^\mu$, and (c) stress tensor $T^{\mu\nu}$ at different system size $N_{mf}$. The horizontal grey bars denote the anticipated integer-spaced level from the conformal symmetry. The rest operators can also be classified into multiplets. (d) The finite-size scaling of the conformal cost function $Q$.}
    \label{fig:3}
\end{figure}

We have checked several criteria for the conformal symmetry from the spectrum. We find a conserved symmetry current $J^\mu$ which has $l=1,s=1$ and scaling dimension $\Delta_J\approx2$, and a $l=2,s=0$ conserved stress tensor $T^{\mu\nu}$ with $\Delta_T\approx3$. Besides that, all the operators can be classified into primaries and their descendants formed by acting spatial derivatives on the primaries and, for spinning primaries, contracting with the fully antisymmetric tensor $\epsilon^{\mu\nu\rho}$~\cite{Zhu2022}. In Fig.~\ref{fig:3}a--c, we show the example of multiplets of the lowest primaries, including the lowest singlet $S$, conserved symmetry current $J^\mu$ and stress tensor $T^{\mu\nu}$. We identify all of their lowest-lying descendants up to $\Delta<6$, and their scaling dimensions are very close to the integer spacing with the primary.

We use a cost function $Q$ to evaluate the quality of the conformal symmetry. It is defined by considering the lowest descendants and conserved currents $(J^\mu,\partial^\mu S,\epsilon^{\mu\nu\rho}\allowbreak\partial_\nu J_\rho,\partial^\mu J^\nu,T^{\mu\nu})$ with $\Delta\lesssim 3$ and taking the root mean square of their deviation from the expected values $(2,\Delta_S+1,3,3,3)$ with exact conformal symmetry. We minimise the cost function to obtain the rescaling factor $v$ and the value $\mu_c$ with best conformal symmetry. We find $\mu_c=0.312$ nearly independent of the system size and take this point as the critical point. The minimum cost function scales to zero at large size (Fig.~\ref{fig:3}d). 

\begin{figure}[t]
    \centering
    \includegraphics[width=\figurewidth]{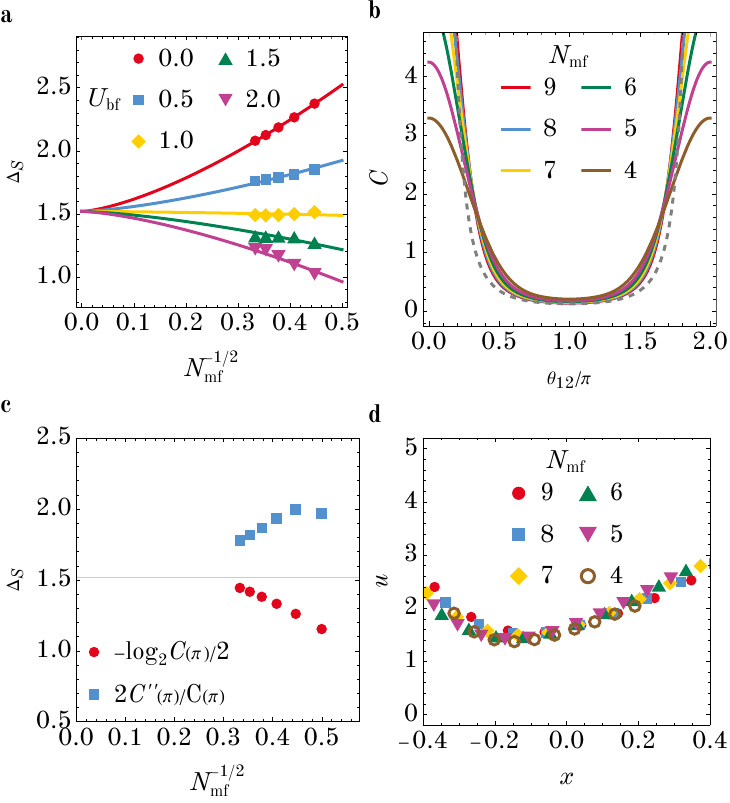}
    \caption{(a) The finite-size scaling of the scaling dimension $\Delta_S$ of the lowest singlet from the spectrum at different $U_{bf}$, indicating that $\Delta_S=1.52(18)$. (b) The two-point correlation function $C(r_{12})=\langle S(\br_1)S(\br_2)\rangle$ as a function of the angular distance $\theta_{12}$ by expressing $S(\br)$ in terms of $n_f(\br)$. (c) The scaling dimensions $\Delta_S$ extracted from the 2-pt function in different ways in Eq.~\eqref{eq:dimless}, compared with the estimate $1.52$. (d) The data collapse of the dimensionless quantity as a function of $x=(\mu-\mu_c)N_{mf}^{(3-\Delta_S)/2}$.}
    \label{fig:4}
\end{figure}

Here we analyse some of the lowest operators. The only relevant primary besides $J^\mu$ is the lowest symmetry-singlet $S$. To estimate its scaling dimension, we need to eliminate the finite-size correction. For that purpose, we tune $U_{bf}$ in the microscopic Hamiltonian, and perform finite-size scaling for different choices of $U_{bf}$. In the field theory description this corresponds to tuning the coefficient of the higher irrelevant singlets. For each $U_{bf}$ and each system size $N_{mf}$, we extract the scaling dimension $\Delta_S$ at their own $\mu_c$ with optimal conformality. We find that $\Delta_S$ scales towards the same direction as $N_{mf}\to\infty$ for different $U_{bf}$ (Fig.~\ref{fig:4}a). We then perform a finite size scaling with the ansatz $\Delta_S(U_{bf},N_{mf})=\Delta_S+g(U_{bf})N_{mf}^{-\omega/2}$. The result is 
\begin{equation*}
    \Delta_S=1.52(18)
\end{equation*}
and the subleading exponent is $\omega=1.4(5)$ (\textit{cf.} Appendix~\ref{app:c}). It is also worth noting that the finite-size correction is the smallest around $U_{bf}=1$.

The source of the correction is higher operators with $s=0$. The lowest ones include a $l=0$ operator at $\Delta=3.7$ and a $l=2$ operator at $\Delta=4.5$ from the $N_{mf}=9$ spectrum. These numbers are subject to finite-size effect. Besides the $s=0$ operators, we observe two lowest $s=2$ primaries. At $N_{mf}=9$, the one with $l=0$ lies around $\Delta=5.1$ and the other with $l=2$ lies around $\Delta=5.3$. The $s^z=\pm 2$ components of these operators correspond to the $4\pi$-monopoles in the $\rU(1)$ gauge theory descriptions. Its scaling dimension extrapolates to $4.888+\mathcal{O}(1/N)$ in large-$N$ expansion~\cite{Chester2022Monopole}. Besides the symmetry current, there exists another $l=1,s=1$ primary that lies around $\Delta=4.5$ at $N_{mf}=9$ (\textit{cf.} Appendix~\ref{app:d}). 
 
Besides the spectrum, we perform further check by studying the local observables, among which the simplest is the fermion density operator $n_f(\br)$. At the conformal point, it is the linear combination of the scaling operators in the scalar representation of $\SO(3)$, \textit{viz.} $\bbI$, $S$, \textit{etc.} ~\cite{Hu2023Mar,Han2023Jun}
\begin{equation}
    n_f(\br)=c_0+c_SS(\br)+\dots
\end{equation}
Using $n_f$, we first calculate the dimensionless two-point correlation function~\cite{Han2023Jun} $C(\theta_{12})=\langle\bbI|S(\br_1)S(\br_2)|\bbI\rangle\allowbreak R^{2\Delta_S}$, where $\theta_{12}$ is the angular distance between the two points (Fig.~\ref{fig:4}b, \textit{cf.} Appendix~\ref{app:e}). The numerical results approach $(2\sin\theta_{12}/2)^{-2\Delta_S}$ required by conformal kinematics. Quantitatively, the scaling dimension can be extracted from the correlator and its derivative at the antipodal points
\begin{equation}
    \Delta_S=\left.-\tfrac{1}{2}\log_2C(\theta_{12})\right|_{\theta_{12}=\pi}=\left.2C''(\theta_{12})/C(\theta_{12})\right|_{\theta_{12}=\pi}.
    \label{eq:dimless}
\end{equation}
The values of $\Delta_S$ extracted in these two ways approaches our previous estimate $1.52(18)$ with increasing system size (Fig.~\ref{fig:4}c). Away from the critical point, these two estimates can be regarded as a dimensionless quantity. We take the average of these two estimates to reduce subleading corrections and denote it as $u(N_{mf},\mu)$. The finite-size scaling hypothesis~\cite{Fisher1972Scaling,Cardy1996Scaling} requires that $u(N_{mf},\mu)=g((\mu-\mu_c)R^{1/\nu})$ where $\nu=1/(3-\Delta_S)$ and $g$ is the scaling function independent of the system size. By plotting $u$ against $x=(\mu-\mu_c)N_{mf}^{(3-\Delta_S)/2}$, curves at different $N_{mf}$ collapses onto each other (Fig.~\ref{fig:4}d), which confirms that the phase transition is driven by the $S$ operator. 

\textit{Discussion} ---
In this paper, we study the Chern-Simons-matter theory of the confinement transition of the KL-CSL on the fuzzy sphere. We study the setup of two flavours of $\SU(2)$-fundamental fermions and one flavour of charge-2 bosons. By tuning a relative chemical potential, we realise a direct transition between a bFQH phase with $\nu_b=1/2$ and a fIQH phase with $\nu_f=2$. We show that this transition can be described by four CS-matter theories that are conjectured to be dual. We present numerical results by exact diagonalisation showing that this transition is continuous and the critical point has emergent conformal symmetry, evidenced by conserved currents, integer-spaced conformal multiplets and a conformal 2-pt correlator. We show that the only relevant primary except symmetry current is a symmetry-singlet and extract its scaling dimension $\Delta_S=1.52(18)$ through a finite-size scaling. We also discuss other primary operators with higher scaling dimensions in this theory. 

This is the first time that this transition has been confirmed to have emergent conformal symmetry and the non-perturbative results for its critical data has been reported. As the only relevant operators are a singlet and a conserved current, in setups with either $\SO(3)$ or only $\rO(2)$ symmetry, the transition can be realised by tuning only one parameter. When the microscopic symmetry is only $\rO(2)$ (or even a discrete subgroup such as $D_4$), the $\SO(3)$ symmetry will still emerge at the transition since perturbations with $s\geq2$ are all irrelevant. The emergent $\SO(3)$ symmetry agrees well with the expectation from the field theory dualities. As the singlet has scaling dimension less than $2$, this transition is unstable against disorder~\cite{Harris1974Disorder}. This CFT has attracted wide attention in various contexts. It describes transitions related to CSL. For example, in the Hofstadter-Hubbard model~\cite{Kuhlenkamp2022CSL,Divic2024CSL}, it describes the transition between IQH insulator and CSL. This model has a pseudospin $\SU(2)_c$ symmetry that contains the charge conservation $\rU(1)$ as its subgroup and becomes gapless at the critical point and a spin $\SU(2)$ symmetry that stays gapped. Our results confirm that at the critical point in this setup, the charge-1 excitation stays gapped and the charge-2 excitation, corresponding to a component of the conserved current $J^\mu$ of the $\SU(2)_c$, becomes gapless. This is crucial for the anyon superconductivity: Upon doping at the critical point, the particles form charge-2 Cooper pairs~\cite{Divic2025AnyonSC}.

The theory we have studied is one of the simplest interacting critical CS-matter conformal field theories. Many of its properties remain to be explored. The scaling dimensions we have obtained may be useful as input in conformal bootstrap~\cite{Poland2018Bootstrap,Rychkov2023Bootstrap} to give more precise estimates. The sparse distribution of primary operators makes this CFT a suitable target for conformal bootstrap, especially for the non-Abelian current bootstrap where the parity breaking can be straightforwardly imposed~\cite{He2023Bootstrap}. Another future direction is to study its defects and boundaries. The $\SU(2)$ gauge theory description admits a Wilson line defect whose endpoints carry a spinor representation of $\SO(3)$. The $\rU(1)$ gauge theory description also admits a Wilson line and a 't Hooft line defect. It would be interesting to realise them on the fuzzy sphere. As a CFT that breaks parity symmetry, its conformal boundary remains largely unknown as well. 

The technique of constructing transitions between different IQH and FQH phases open up new possibilities to study various other critical gauge theories. For example, the transition between CSL and a superfluid with $\rU(1)$ symmetry-breaking or CSL and N\'eel order with $\SU(2)$ symmetry-breaking, described by $\rU(1)_2$ with 2 scalars and its easy-plane version~\cite{Song2022CSL}, can be constructed in a similar manner. Some of the candidate CS-matter theories for the $\Sp(N)$ CFTs~\cite{Zhou2024Oct} can be constructed explicitly. Similar setup may also be used to address some of the outstanding problems in condensed matter physics, \textit{e.g.}, the nature of Dirac spin liquid. Furthermore, using constructions with both bosons and fermions, it is promising to realise CFTs with fermionic operator content.

\textit{Acknowledgments} ---
We would like to thank Shai Chester, Zixiang Hu, Ryan Lanzetta, Max Metlitski, and Ashvin Vishwanath for illuminating discussions. Z.Z. acknowledges support from the Natural Sciences and Engineering Research Council of Canada (NSERC) through Discovery Grants. Research at Perimeter Institute is supported in part by the Government of Canada through the Department of Innovation, Science and Industry Canada and by the Province of Ontario through the Ministry of Colleges and Universities.

\let\oldaddcontentsline\addcontentsline
\renewcommand{\addcontentsline}[3]{}

\let\addcontentsline\oldaddcontentsline

\onecolumngrid
\tableofcontents
\appendix

\section{The CS-matter theory descriptions of the fIQH-bFQH transition}
\label{app:a}

In this section, we discuss the four Chern-Simons matter theories that describe the transition between $\nu_f=2$ fIQH and $\nu_b=1/2$ bFQH, especially how they can produce the correct response~\cite{Hsin2016LevelRank,Benini2017Duality,Seiberg2016Duality}. 

We start with a single flavour of critical complex boson coupled to $\rU(1)_{2}$ Chern-Simons theory
\begin{equation}
    \cL_1=|(\partial_\mu-ib_\mu+iA_\mu)\phi|^2+\frac{2}{4\pi}b\,\rd b-m^2|\phi|^2-\lambda|\phi|^4+2\CS_g
\end{equation}
Here we denote the dynamical gauge field by $b$. The Lagrangian only have a manifest $\rO(2)$ global symmetry. The $\rU(1)$ part is the flux conservation of the gauge field, and $A_\mu$ is the probe field for this symmetry; the improper $\bbZ_2$ is the charge conjugation. The gravitational Chern-Simons term is defined through extending into a 4d manifold $X$ with the original 3d manifold its boundary $M=\partial X$~\cite{Hsin2016LevelRank}
\begin{equation*}
    \int_{M=\partial X}\CS_g=\frac{1}{192\pi}\int_X\tr R\wedge R
\end{equation*}
where $R$ is the curvature 2-form. The choice of the counterterms ensures the correct responses on the two sides of the transition. This theory can equivalently be written as 
\begin{equation}
    \cL_1=|(\partial_\mu-ib'_\mu)\phi|^2+\frac{2}{4\pi}b'\rd b'+\frac{2}{2\pi}b'\rd A+\frac{2}{4\pi}A\,\rd A-m^2|\phi|^2-\lambda|\phi|^4+2\CS_g
\end{equation}
where $b'=b-A$, from which it is manifest that $A$ is coupled to the symmetry current current $J_{\text{top}}=\star\,\rd a$ of the topological $\rU(1)$ symmetry. The phase transition is controlled by $m^2$. Positive $m^2$ gaps the scalar field and gives a $\rU(1)_{2}$ of the dynamical gauge field $b$ with response $\kappa_{xy}=1,\sigma_{xy}=0$. It is topologically ordered due to the non-trivial Chern-Simons term. Acute readers may notice that this CS theory describes anti-semion topological order, rather than the semion topological order of the $\nu = 1/2$ Laughlin state. Nevertheless, our system contains local fermions, so one can always attach a fermion to an anyon, transforming a semion into an anti-semion and vice versa. In other words, in the presence of local fermions, semion topological order is equivalent to anti-semion topological order; strictly speaking, it is a spin-TQFT.
 Negative $m^2$ condenses the scalar field and Higgses the gauge field imposing $b=A$. This gives a trivially gapped phase with response $\kappa_{xy}=2,\sigma_{xy}^{\rU(1)\subset\SU(2)}=2$.
\begin{align}
    \cL_1^>&=\frac{2}{4\pi}b\,\rd b+2\CS_g\xrightarrow{\int\mathcal{D}b}\CS_g\\
    \cL_2^<&=\frac{2}{4\pi}A\,\rd A+2\CS_g.
\end{align}

We then analyse a single flavour of critical Dirac fermion coupled to $\rU(1)_{-3/2}$ Chern-Simons theory, whose Lagrangian is 
\begin{equation}
    \cL_2=\bar{\psi}(i\slashed{\partial}+\slashed{a}-m)\psi-\frac{3}{8\pi}a\,\rd a-\frac{2}{2\pi}A\,\rd a-\frac{2}{4\pi}A\,\rd A+\frac{1}{2}\CS_g
\end{equation}
Here $a_\mu$ is a $\rU(1)$ dynamical gauge field.
The mass $m$ is a parameter that controls the phase transition. At critical $m=m_c$ (which is $0$ semi-classically), this Lagrangian flows to a $\SO(3)$-symmetric CFT, and the manifest $\rU(1)$ is its subgroup. Away from the critical point, a gap opens and the Dirac fermion induces a Chern-Simons response of the gauge field $a$ with level $\tfrac{1}{2}\sgn(m-m_c)$ and a gravitational Chern-Simons response with the same level~\cite{Lee2018QEDCS,Ma2020QCD}
\begin{align}
    \cL_2^<&=-\frac{2}{4\pi}a\,\rd a-\frac{2}{2\pi}A\,\rd a-\frac{2}{4\pi}A\,\rd A\nonumber\\
    &=-\frac{2}{4\pi}(a+A)\rd(a+A)\xrightarrow{\int\mathcal{D}a}\CS_g\\
    \cL_2^>&=-\frac{1}{4\pi}a\,\rd a-\frac{2}{2\pi}A\,\rd a-\frac{2}{4\pi}A\,\rd A+\CS_g\nonumber\\
    &=-\frac{1}{4\pi}(a+2A)\rd(a+2A)+\frac{2}{4\pi}A\,\rd A+\CS_g\xrightarrow{\int\mathcal{D}a}\frac{2}{4\pi}A\,\rd A+2\CS_g
\end{align}
At $m<m_c$, the dynamical $\rU(1)$ gauge field has a Chern-Simons level $-2$, corresponding to the semion topological order; at $m>m_c$, it has level $-1$, corresponding a non-topologically ordered gapped phase. To obtain the response to the external field, we integrate out the dynamical gauge field (denoted by $\int\mathcal{D}a$) to obtain a Lagrangian depending on only the probe field $A$ and the curvature $R$. Integrating out the $\rU(1)$ gauge field produces a gravitational Chern-Simons level-$\pm 1$ whose sign is opposite to the level of $a$. The Chern-Simons level of the probe field $A$ gives the $\rU(1)$ Hall conductance in the unit of $e^2/h$, and the gravitational Chern-Simons level gives the thermal Hall conductance $\kappa_{xy}$ in the unit of $(\pi/6)(k_B^2T/\hbar)$. Hence, at $m<m_c$, the response is $\sigma_{xy}=0$ and $\kappa_{xy}=1$, which is consistent with the $\nu_b=1/2$ bFQH phase; at $m>m_c$, the response is $\sigma_{xy}^{\rU(1)\subset\SU(2)}=2$ and $\kappa_{xy}=2$, which is consistent with the $\nu_f=2$ fIQH phase. We will show later that $\sigma_{xy}^{\rU(1)\subset\SU(2)}=2$ is equivalent to $\sigma_{xy}^{\SU(2)}=1$.

We then move on to the $\SU(2)$ gauge theories. The fermionic version is a critical Dirac fermion coupled to $\SU(2)_1$ Chern-Simons theory~\cite{Ma2020QCD}
\begin{equation}
    \cL_3=\bar{\Psi}(i\slashed{\partial}+\slashed{\alpha}+\slashed{B}-m)\Psi-\frac{1}{8\pi}\CS[\alpha]+\frac{1}{8\pi}\CS[B]+\CS_g
\end{equation}
where $\Psi$ and $\Phi$ are $\SU(2)$-fundamental Dirac fermion and complex scalar, $\alpha_\mu$ and $\beta_\mu$ are dynamical $\SU(2)$ gauge fields and $B_\mu$ is a $\SU(2)$ probing field. The shorthand notation 
\begin{equation*}
    \CS[\alpha]=\epsilon^{\mu\nu\rho}\tr\left(\alpha_\mu\partial_\nu\alpha_\rho+\frac{2}{3}\alpha_\mu\alpha_\nu\alpha_\rho\right)
\end{equation*}
Note that here the dynamical gauge field $\alpha$ couples to the gauge degree of freedom, while the probe field $B$ couples to the flavour degree of freedom, \textit{i.e.}, $\alpha$ is an adjoint under gauge $\SU(2)$ and a singlet under flavour $\SU(2)$, while $B$ is a singlet under gauge $\SU(2)$ and an adjoint under flavour $\SU(2)$. To be more explicit, we expand the components in the expression. 
\begin{gather}
    \bar{\Psi}(i\slashed{\partial}+\slashed{\alpha}+\slashed{B})\Psi=\bar{\Psi}_{fia}i(\gamma^\mu)^a{}_b(D_{\alpha,B,\mu})^{fi}{}_{gj}\Psi^{gbj}\\
    (D_{\alpha,B,\mu})^{fi}{}_{gj}=\delta^f_g\delta^i_j\partial_\mu-i\delta^f_g(\alpha_\mu)^i{}_j-i\delta^i_j(B_\mu)^f{}_g,
\end{gather}
where we use $f,g,h$ as the $\SU(2)$ flavour indices, $i,j,k$ as the $\SU(2)$ gauge indices, $a,b$ as the $\SO(3)$ spinor indices, and the adjoint is defined as $\bar{\Psi}_{fia}=\Psi^{gjb}\epsilon_{fg}\epsilon_{ij}\epsilon_{ab}$. Finite $m$ opens a gap for the Dirac fermion and induces a Chern-Simons response of the dynamic gauge field $\alpha$, probe field $B$ and gravity with level $\tfrac{1}{2}\sgn m$. Similar to the analyses above, 
\begin{align}
    \cL_3^<&=-\frac{1}{4\pi}\CS[\alpha]\xrightarrow{\int\mathcal{D}\alpha}\CS_g\\
    \cL_3^>&=\frac{1}{4\pi}\CS[B]+2\CS_g.
\end{align}
One side of the critical point flows to the $\SU(2)_1$ Chern-Simons theory with semion topological order and response $\kappa_{xy}=1,\sigma^{\SU(2)}_{xy}=0$; the other side flows to a trivially gapped phase with response $\kappa_{xy}=2,\sigma^{\SU(2)}_{xy}=1$. Note that $(1/4\pi)\CS[B]$ and $(2/4\pi)A\,\rd A$ give the same response. To see this, let $B=A\sigma^z$,
\begin{equation}
    \frac{1}{4\pi}\CS[B]=\frac{1}{4\pi}\tr\left((A\sigma^z)\,\rd (A\sigma^z)+\frac{2}{3}(A\sigma^z)^3\right)=\frac{2}{4\pi}A\,\rd A.
\end{equation}

Lastly, for a single flavour of critical complex boson coupled to $\SU(2)_1$ Chern-Simons theory
\begin{equation}
    \cL_4=|(\partial_\mu-i\beta_\mu-iB_\mu)\Phi|^2+\frac{1}{4\pi}\CS[\beta]-m^2|\Phi|^2-\lambda|\Phi|^4+2\CS_g
\end{equation}
More explicitly, 
\begin{equation}
    |(\partial_\mu-i\beta_\mu-iB_\mu)\Phi|^2=-\bar{\Phi}_{fi}(D_{\beta,B,\mu})^{fi}{}_{gj}(D_{\beta,B}{}^\mu)^{gj}{}_{hk}\Phi^{hk},
\end{equation}
where $\bar{\Phi}_{fi}=\Phi^{gj}\epsilon_{fg}\epsilon_{ij}$. Positive $m^2$ gaps the scalar and negative $m^2$ Higgses the gauge field.
\begin{align}
    \cL_4^>&=\frac{1}{4\pi}\CS[\beta]+2\CS_g\xrightarrow{\int\mathcal{D}\beta}\CS_g\\
    \cL_4^<&=\frac{1}{4\pi}\CS[B]+2\CS_g
\end{align}

\section{Technical details of the ED calculation}
\label{app:b}

In this section, we present the technical details of the ED calculation. More details can be found in the documentation of the FuzzifiED package~\cite{FuzzifiED}. A sample code can be found in the GitHub repository at \href{https://github.com/FuzzifiED/FuzzifiED.jl/blob/main/examples/su2_1_scal_spectrum.jl}{FuzzifiED.jl/examples/su2\_1\_scal\_spectrum.jl}.

The $\rU(1)$ and $\mathbb{Z}_2$ subgroups of the rotation and global symmetry are implemented to divide the Hilbert space into sectors. We implement the conservation of electric charge $q_e$, one component of angular momentum $L^z$ and spin $S^z$ of global symmetry
\begin{align}
    N_e&=\sum_mf^\dagger_{mi}f_m^i+\sum_m2b_m^\dagger b_m\nonumber\\
    L_z&=\sum_mmf^\dagger_{mi}f_m^i+\sum_mmb_m^\dagger b_m\nonumber\\
    S_z&=\sum_mf_{m1}^\dagger f_m^1-f_{m2}^\dagger f_m^2.
\end{align}
To go through all the symmetry representations, it suffices to consider only the $L_z=0,S_z=0$ sector. This sector has two additional $\bbZ_2$ symmetries
\begin{align}
    \mathcal{R}_y:f_m^a&\mapsto(-1)^{s-m}f_{-m}^a,\quad b_m\mapsto(-1)^{2s-m}b_{-m}\nonumber\\
    \mathcal{Z}:f_m^1&\mapsto f_m^2,\quad f_m^2\mapsto-f_m^1
\end{align}
corresponding to the $\pi$-rotation along the $y$-axis and the exchange of two flavours. States with even $l$ are even under $\mathcal{R}_y$ and vice versa; states with even $s$ are even under $\mathcal{Z}$ and vice versa.

To determine the representations of each state under $\SO(3)$ rotation and global symmetry, we measure the total angular momentum $L^2$ and the total spin $S^2$. The $L^2$ operator can be constructed using $L_{z,\pm}$ in the following way
\begin{align}
    L_z&=\sum_mmf^\dagger_{mi}f_m^i+\sum_mmb_m^\dagger b_m\nonumber\\
    L_\pm&=\sum_{ma}\sqrt{(s\mp m)(s\pm m+1)}f^\dagger_{m\pm1,i}f_m^i+\sum_{ma}\sqrt{(2s\mp m)(2s\pm m+1)}b^\dagger_{m\pm1}b_m\nonumber\\
    L^2&=L_+L_-+L_z^2-L_z.
\end{align}
Its expectation value is related to the rotation $\SO(3)$ spin of the state by $\langle\Phi|L^2|\Phi\rangle=l_\Phi(l_\Phi+1)$
The total spin $S^2$ of the global $\SO(3)$ symmetry is 
\begin{align}
    S^2=\left[\frac{1}{2}\sum_m f_{mi}^\dagger(\sigma^\mu)^i{}_jf_m^j\right]^2
\end{align}
Its expectation value is related to the global $\SO(3)$ spin of the state by $\langle\Phi|S^2|\Phi\rangle=s_\Phi(s_\Phi+1)$.

\section{Estimation of singlet scaling dimension}
\label{app:c}

In this section, we present the numerical details for estimating the scaling dimension $\Delta_S$ of the relevant singlet. 

\begin{figure}[htbp]
    \centering
    \includegraphics[width=0.5\linewidth]{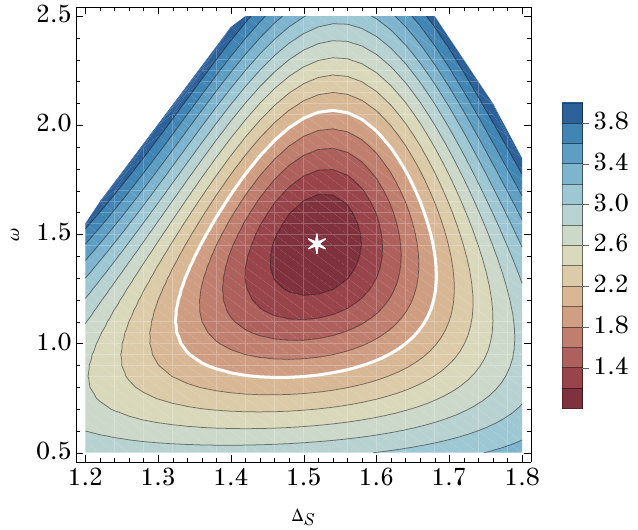}
    \caption{The rescaled fitting residual $\sigma/\sigma_{\min}$ as a function of parameters $\Delta_S$ and $\omega$. The white line denotes the contour $\sigma/\sigma_{\min}=2$ that we take as the confidence region and the star denotes the optimal point.}
    \label{fig:c}
\end{figure}

The $\Delta_S$ and $\omega$ are estimated from $\Delta_S^{(\text{FS})}$ measured from the spectrum at different $U_{bf},N_{mf}$. We fit with the ansatz 
\begin{equation}
    \tilde{\Delta}_S(U_{bf},N_{mf};\Delta_S,\omega,g(U_{bf}))=\Delta_S+g(U_{bf})N_{mf}^{-\omega/2}
\end{equation}
and minimise 
\begin{equation}
    \sigma^2(\Delta_S,\omega,g(U_{bf}))=\sum_{\{U_{bf},N_{mf}\}}\left[\Delta_S^{(\text{FS})}(U_{bf},N_{mf})-\tilde{\Delta}_S(U_{bf},N_{mf};\Delta_S,\omega,g(U_{bf}))\right]^2
\end{equation}
with respect to $\Delta_S,\omega$ and $g(U_{bf})$. In Fig.~\ref{fig:c}, we show $\sigma$ as a function of $\Delta_S$ and $\omega$, \textit{i.e.}, we fix $\Delta_S$ and $\omega$, and for each $\Delta_S$ and $\omega$, we minimise $\sigma$ with respect to $g(U_{bf})$. We take the region with $\sigma/\sigma_{\min}<2$ as the confidence region. The estimation is 
\begin{equation}
    \Delta=1.52(18),\qquad \omega=4.4(5).
\end{equation}

\section{Full spectrum}
\label{app:d}

\begin{figure}[htbp]
    \centering
    \includegraphics[width=0.6\linewidth]{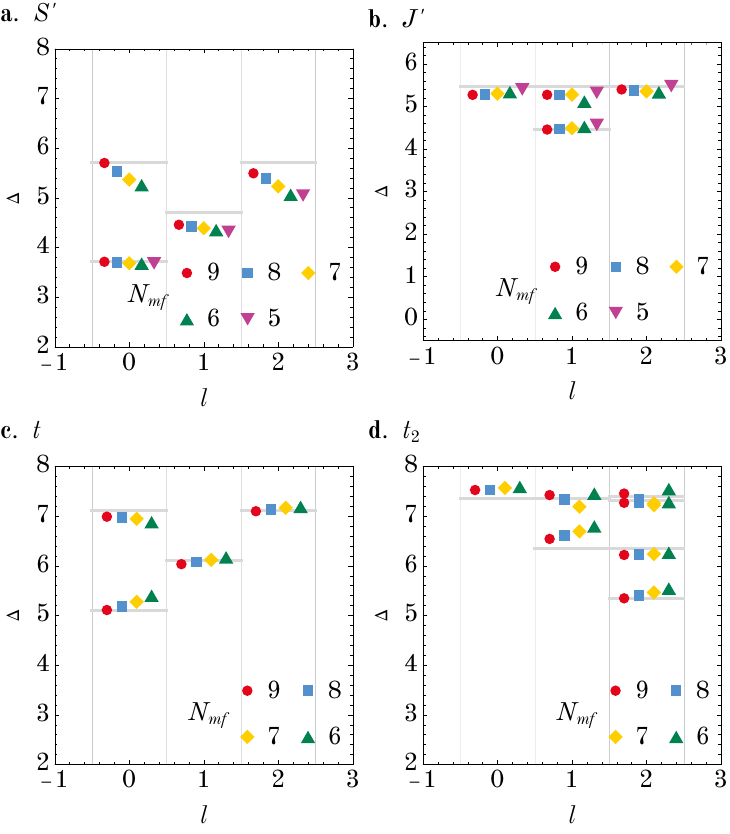}
    \caption{The scaling dimensions of the conformal multiplet at different system size $N_{mf}$ of (a) the second lowest singlet $S'$, (b) the second lowest $s=1,l=1$ operator $J'^\mu$, (c) the lowest $s=2,l=0$ operator $t$, and (d) the lowest $s=2,l=2$ operator $t_2^{\mu\nu}$. The horizontal grey bars denote the anticipated integer-spaced level from the conformal symmetry, where the scaling dimensions are taken from $N_{mf}=9$ raw data.}
    \label{fig:d}
\end{figure}

In Tables~\ref{tbl:spec0}, \ref{tbl:spec1} and \ref{tbl:spec2}, we list the spectrum measured at $\mu=0.312$ and $N_{mf}=9$ with $l\leq 2$ and $\Delta<6$. For the $s=2$ sector, we calculate up to $\Delta\lesssim 7.5$ to target more descendants. We identify the multiplets of $S,S',J^\mu,J'^\mu,T^{\mu\nu},T'^{\mu\nu},t$ and $t_2^{\mu\nu}$ mentioned in the main text. This include all the operators up to $\Delta<4.7$. The rest unidentified operators may belong to multiplets of higher primaries. We plot the conformal multiplets of $S',J'^\mu,t$ and $t_2^{\mu\nu}$ in Fig.~\ref{fig:d}. 

\begin{table}[htbp]
    \centering
    \begin{tabular}{cccc|cccc|cccc}
        \hline\hline
        $l$ & $s$ & $\Delta$ & op. & $l$ & $s$ & $\Delta$ & op. & $l$ & $s$ & $\Delta$ & op.\\
        \hline
        0 & 0 & 0.0000 & $\bbI$ & 1 & 0 & 2.3385 & $\partial^\mu S$ & 2 & 0 & 2.9595 & $T^{\mu\nu}$ \\
        0 & 0 & 1.4895 & $S$ & 1 & 0 & 4.0081 & $\Box\partial^\mu S$ & 2 & 0 & 3.0828 & $\partial^\mu\partial^\nu S$\\
        0 & 0 & 3.1813 & $\Box S$ & 1 & 0 & 4.4552 & $\partial^\mu S'$ & 2 & 0 & 3.9347 & $\epsilon^{\mu\rho}{}_\sigma\partial_\rho T^{\nu\sigma}$\\\
        0 & 0 & 3.7102 & $S'$ & 1 & 0 & 5.5475 & $\partial_\nu T'^{\mu\nu}$ & 2 & 0 & 4.4782 & $T'^{\mu\nu}$\\
        0 & 0 & 4.9903 & & 1 & 0 & 5.6311 & & 2 & 0 & 4.7108\\
        0 & 0 & 5.2927 & $\Box^2 S$ & 1 & 0 & 5.7842 & & 2 & 0 & 4.7131\\
        0 & 0 & 5.6991 & $\Box S'$ & 1 & 0 & 5.8546 & & 2 & 0 & 4.9981\\
        0 & 0 & 5.8729 & & 1 & 0 & 5.9726 & & 2 & 0 & 5.0312 & $\Box T^{\mu\nu}$\\
        0 & 0 & 5.9171 & & & & & & 2 & 0 & 5.2351 & $\Box\partial^\mu\partial^\nu S$ \\
        & & & & & & & & 2 & 0 & 5.4913 & $\partial^\mu\partial^\nu S'$\\
        & & & & & & & & 2 & 0 & 5.5123 & $\epsilon^{\mu\rho}{}_\sigma\partial_\rho T'^{\nu\sigma}$ \\
        & & & & & & & & 2 & 0 & 5.7519\\
        & & & & & & & & 2 & 0 & 5.7776\\
        & & & & & & & & 2 & 0 & 5.8263\\
        & & & & & & & & 2 & 0 & 5.8610\\
        & & & & & & & & 2 & 0 & 5.9085\\
        & & & & & & & & 2 & 0 & 5.9343\\
        & & & & & & & & 2 & 0 & 5.9612\\
        \hline\hline
    \end{tabular}
    \caption{The scaling dimensions of the operators with $l\leq 2$, $s=0$ and $\Delta<6$ at $\mu=0.312$ and $N_{mf}=9$.}
    \label{tbl:spec0}
\end{table}

\begin{table}[htbp]
    \centering
    \begin{tabular}{cccc|cccc|cccc}
        \hline\hline
        $l$ & $s$ & $\Delta$ & op. & $l$ & $s$ & $\Delta$ & op. & $l$ & $s$ & $\Delta$ & op.\\
        \hline
        0 & 1 & 5.2687 & $\partial_\mu J'^\mu$ & 1 & 1 & 2.0300 & $J^\mu$ & 2 & 1 & 2.9582 & $\partial^\mu J^\nu$\\
        & & & & 1 & 1 & 3.1001 & $\epsilon^{\mu\nu\rho}\partial_\nu J_\rho$ & 2 & 1 & 3.8893 & $\epsilon^{\mu\sigma\rho}\partial^\nu\partial_\sigma J_\rho$\\
        & & & & 1 & 1 & 4.1108 & $\Box J^\mu$ & 2 & 1 & 4.5707\\
        & & & & 1 & 1 & 4.4540 & $J'^\mu$ & 2 & 1 & 4.7688 & $\Box\partial^\mu J^\nu$\\
        & & & & 1 & 1 & 4.9796 & $\Box\epsilon^{\mu\nu\rho}\partial_\nu J_\rho$ & 2 & 1 & 5.2225\\
        & & & & 1 & 1 & 5.2702 & & 2 & 1 & 5.3703\\
        & & & & 1 & 1 & 5.4766 & $\epsilon^{\mu\nu\rho}\partial_\nu J'_\rho$ & 2 & 1 & 5.3957 & $\partial^\mu J'^\nu$\\
        & & & & 1 & 1 & 5.6241 & & 2 & 1 & 5.7839\\
        & & & & 1 & 1 & 5.6986 & & 2 & 1 & 5.8092\\
        & & & & 1 & 1 & 5.9216 & & 2 & 1 & 5.8573\\
        & & & & 1 & 1 & 5.9495 & & 2 & 1 & 5.9906\\
        & & & & 1 & 1 & 5.9869 & & 2 & 1 & 5.9979\\
        \hline\hline
    \end{tabular}
    \caption{The scaling dimensions of the operators with $l\leq 2$, $s=1$ and $\Delta<6$ at $\mu=0.312$ and $N_{mf}=9$.}
    \label{tbl:spec1}
\end{table}

\begin{table}[htbp]
    \centering
    \begin{tabular}{cccc|cccc|cccc}
        \hline\hline
        $l$ & $s$ & $\Delta$ & op. & $l$ & $s$ & $\Delta$ & op. & $l$ & $s$ & $\Delta$ & op.\\
        \hline
        0 & 2 & 5.1079 & $t$ & 1 & 2 & 6.0314 & $\partial^\mu t$ & 2 & 2 & 5.3435 & $t_2^{\mu\nu}$ \\
        0 & 2 & 6.9871 & $\Box t$ & 1 & 2 & 6.5398 & $\partial_\nu t_2^{\mu\nu}$ & 2 & 2 & 6.2170 & $\epsilon^{\mu\rho}{}_\sigma\partial_\rho t_2^{\nu\sigma}$ \\
        0 & 2 & 7.5240 & $\partial_\mu\partial_\nu t_2^{\mu\nu}$ & 1 & 2 & 7.4221 & $\epsilon^{\mu\nu\rho}\partial_\nu\partial^\sigma t_{2,\rho\sigma}$ & 2 & 2 & 6.6462 \\
        & & & & & & & & 2 & 2 & 6.8537 \\
        & & & & & & & & 2 & 2 & 7.0139 \\
        & & & & & & & & 2 & 2 & 7.0966 & $\partial^\mu\partial^\nu t$ \\
        & & & & & & & & 2 & 2 & 7.2665 & $\Box t_2^{\mu\nu}$ \\
        & & & & & & & & 2 & 2 & 7.4528 & $\partial^\mu\partial_\rho t_2^{\nu\rho}$ \\
        \hline\hline
    \end{tabular}
    \caption{The scaling dimensions of the operators with $l\leq 2$, $s=0$ and $\Delta\lesssim 7.5$ at $\mu=0.312$ and $N_{mf}=9$.}
    \label{tbl:spec2}
\end{table}

\section{Correlation functions}
\label{app:e}

In this section, we present the detail for calculating the correlation function. 

The fermion density $n_f(\br)=f^\dagger_i(\br)f^i(\br)$ is the linear combination of all the scaling operators with $s=0$~\cite{Hu2023Mar}. In the leading order, it can be used as UV realisations of CFT operators $S$. 
\begin{equation}
    n_f(\br)=\lambda_0+\lambda_SS(\br)+\lambda_{\partial^\mu S}\partial^\mu S(\br)+\dots,\qquad S^{(\text{FS})}=\frac{n_f-\lambda_0}{\lambda_S}.
\end{equation}
The operators on the right-hand side are defined on the cylinder, it is converted with operators on the flat spacetime by 
\begin{equation}
    S^{(\text{cyc.})}(\br,\tau)=R^{-\Delta_S}S^{(\text{flat})}(x),\quad x\in\mathbb{R}^3.
\end{equation}
The coefficients $\lambda_0,\lambda_S,\dots$ can be determined by measuring its matrix elements
\begin{equation}
    \lambda_0=\frac{1}{\sqrt{4\pi}}\langle 0|n_{f,00}|0\rangle,\quad\lambda_S=\frac{R^{\Delta_S}}{\sqrt{4\pi}}\langle S|n_{f,00}|0\rangle
\end{equation}
where the angular components of the density operator is defined as 
\begin{equation}
    n_f(\br)=\sum_{lm}Y_{lm}(\br)n_{lm},\quad n_{f,lm}=\frac{1}{R^2}\int\rd^2\br\,\bar{Y}_{lm}(\br)n_f(\br)
\end{equation}

We consider the correlation function of two points on the unit sphere in the flat spacetime with the angular distance $\theta_{12}$~\cite{Han2023Jun}
\begin{equation}
    C(\theta_{12})=\langle S^{(\mathrm{flat})}(n_1)S^{(\mathrm{flat})}(n_2)\rangle=\left(2\sin\frac{\theta_{12}}{2}\right)^{-2\Delta_S}
\end{equation}
Converted to the cylinder 
\begin{align}
    C(\theta_{12})&=R^{2\Delta_S}\langle0|S(\br_1)S(\br_2)|0\rangle\nonumber\\
    &=R^{2\Delta_S}\frac{\langle 0|n_f(\br_1)n_f(\br_2)|0 \rangle-\lambda_0^2}{\lambda_S^2}\nonumber\\
    &=\frac{\langle 0|n_f(\br_1)n_f(\br_2)|0 \rangle-\frac{1}{4\pi}\langle 0|n_{f,00}|0\rangle^2}{\frac{1}{4\pi}\langle S|n_{f,00}|0\rangle^2}
\end{align}
where $\br_1=Rn_1,\br_2=Rn_2$, and the correlation function of the density operator can be expressed in terms of the angular components
\begin{align}
    \langle 0|n_f(\br_1)n_f(\br_2)|0\rangle&=\sum_{lm_1m_2}\langle n_{f,lm_1}n_{f,lm_2}\rangle Y_{lm_1}(\br_1)Y_{lm_2}(\br_2)\nonumber\\
    &=\sum_l(2l+1)P_l(\cos\theta_{12})\langle n_{f,l0}n_{f,l0}\rangle
\end{align}

\begin{figure}[htbp]
    \centering
    \includegraphics[width=0.6\linewidth]{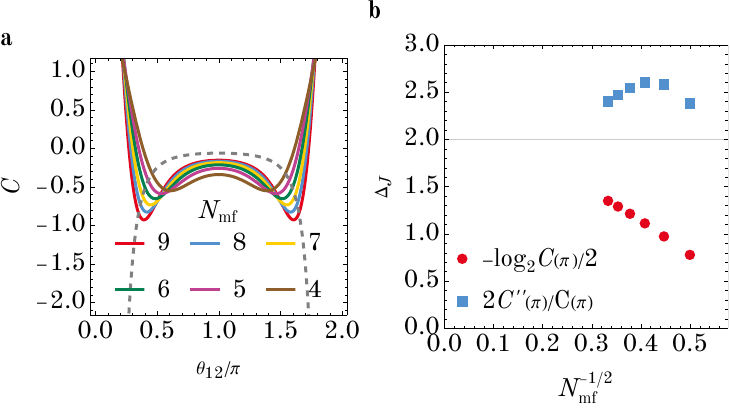}
    \caption{(a) The two-point correlation function $C(r_{12})=\langle J^\tau(\br_1)J^\tau(\br_2)\rangle$ as a function of the angular distance $\theta_{12}$ by expressing $J^\tau(\br)$ in terms of $n^z(\br)$. (c) The scaling dimensions $\Delta_J$ extracted from the 2-pt function in different ways in Eq.~\eqref{eq:dimless}, compared with $2$.}
    \label{fig:e}
\end{figure}

We complement the data by the correlation function of the time component of the symmetry current $J^\tau$ in the leading order, it can be realised by the fermion density operator $n^z$
\begin{equation}
    n^z(\br)=f^\dagger_i(\br)(\sigma^z)^i{}_jf^j(\br)=\lambda_JJ^\tau(\br)
\end{equation}
Note that in Euclidean signature, $J^\tau$ is anti-Hermitian as Hermitian conjugaison acts as time-reflection $(J^\tau)^\dagger=-J^\tau$. Hence
\begin{equation}
    (J^\tau)^{(\text{FS})}=\frac{n^z}{\lambda_J},\qquad \lambda_J=R^2\sqrt\frac{3}{4\pi}\langle J|n^z_{10}|0\rangle
\end{equation}
Note that here we normalise $J^\mu$ by the 2-pt function $\langle J^\mu J^\nu\rangle=x^{-4}(\delta^{\mu\nu}-2x^\mu x^\nu/x^2)$. So we calculate the correlation function 
\begin{equation}
    C(\theta_{12})=R^4\langle0|J^\tau(\br_1)J^\tau(\br_2)|0\rangle=\frac{\langle 0|n^z(\br_1)n^z(\br_2)|0\rangle}{\frac{3}{4\pi}\langle J|n^{z}_{10}|0\rangle^2}.
\end{equation}
We calculate the correlation function, compare it with the conformal correlator $-(2\sin\theta_{12}/2)^{-4}$ and extract the scaling dimension from the correlator~(Fig.~\ref{fig:e}).

\section{Conformal generator}
\label{app:f}

In this section, we present further numerical evidence for the conformal symmetry by studying the conformal generator.

The conformal generator $\Lambda^\mu=P^\mu+K^\mu$ on the states is the $l=1$ component of the Hamiltonian density, which is the local density-density interactions with some full derivatives~\cite{Fardelli2024}
\begin{multline}
    \mathcal{H}(\br)=U_bn_b^2+U_fn_f^2+U_{bf}n_bn_f+t(b^\dagger f^1f^2+\hc)+\mu_fn_f+\mu_bn_b\\+\#\nabla^2n_b+\#\nabla^2(n_b^2)+\#\nabla^2n_f+\#\nabla^2(n_f^2)+\dots
\end{multline}
We have only listed a few examples of the allowed full derivatives. These terms do not affect the Hamiltonian $H=\int\rd^2\br\,\mathcal{H}$. The generator is expressed as 
\begin{equation}
    \Lambda_m=P_m+K_m=\int\rd^2\br\,Y_{1m}(\br)\mathcal{H}(\br).
\end{equation}
To determine those constants, we consider another strategy by considering the $s=0,l=1$ combinations of the fermion and boson operators. Here we include
\begin{multline}
    \Lambda_m=\tilde{U}_f\Delta^\dagger_{f,2s,m_1}\Delta_{f,2s,m_2}\langle2s(-m_1),2sm_2|1m\rangle\\+\tilde{U}_b\Delta^\dagger_{b,4s,m_1}\Delta_{b,4s,m_2}\langle4s(-m_1),4sm_2|1m\rangle+\tilde{U}_{bf}\Delta^\dagger_{bf,i,3s,m_1}\Delta_{bf,3s,m_2}^i\langle3s(-m_1),3sm_2|1m\rangle\\
    +\tilde{t}\left( b^\dagger_{m_1}\Delta_{f,2s,m_2}\langle2s(-m_1),2sm_2|1m\rangle+\hc\right)+\tilde{\mu}f^\dagger_{i m_1}f^i_{m_2}\langle s(-m_1),sm_2|1m\rangle
\end{multline}
where the pairing operators are 
\begin{align}
    \Delta_{f,lm} &=f^1_{m_1}f^2_{m_2}\langle sm_1,sm_2|lm\rangle\nonumber\\
    \Delta_{b,lm} &=b_{m_1}b_{m_2}\langle 2sm_1,2sm_2|lm\rangle\nonumber\\
    \Delta_{bf,lm}^i&=b_{m_1}f^i_{m_2}\langle 2sm_1,sm_2|lm\rangle
\end{align}

In practice, we determine the coefficients by maximising the overlap between $\Lambda^z|S\rangle$ and $\partial_z S$
\begin{equation*}
    \frac{|\langle\partial_z S|\Lambda^z|S\rangle|^2}{\|\Lambda^z|S\rangle\|^2}
\end{equation*}
It is worth noting that $\tilde{U}_b=0$ in the optimal coefficients.

For a state $|\partial^n\Phi,l\rangle$ in the multiplet of primary $\Phi$, the action of $\Lambda^z$ brings it to the superposition of $|\partial^{n\pm 1}\Phi, l'=l\pm 1\rangle$, we decompose $\Lambda^z|\partial^n\Phi,l\rangle$ into angular momentum sectors and check the overlap 
\begin{equation}
    \frac{\left|\langle\partial^{n-1}\Phi, l'|P_{l'}\Lambda_z|\partial^n\Phi,l\rangle\right|^2+\left|\langle\partial^{n+1}\Phi, l'|P_{l'}\Lambda_z|\partial^n\Phi,l\rangle\right|^2}{\left\|P_{l'}\Lambda_z|\partial^n\Phi,l\rangle\right\|^2}
\end{equation}
where $P_{l'}$ means projection to angular momentum $l'$ sector. The results are shown in Tables~\ref{tbl:gen}. Closer total squared overlap to unity exhibits better conformal symmetry. Through these results, we make sure that we have found the correct descendants.

\begin{table}[htbp]
    \centering
    \caption{The overlaps $\langle\partial^{n'}\Phi, l'|P_{l'}\Lambda_z|\partial^n\Phi,l\rangle/\left\|P_{l'}\Lambda_z|\partial^n\Phi,l\rangle\right\|$. The asterisk marks that we have removed the level mixing of the operator. The data are measured at $N_{mf}=7,\mu=\mu_c=0.312$.}
    \label{tbl:gen}
    \begin{tabular}{cc|cc|cc|c}
        \hline\hline
        $|\partial^n\Phi,l\rangle$&$l'$&$\langle\partial^{n'}\Phi,l'|$&$|\textrm{overlap}|^2$&$\langle\partial^{n'}\Phi,l'|$&$|\textrm{overlap}|^2$&total\\
        \hline
        $S$ & $1$ & $\partial^\mu S$ & $0.9990$ & & & $0.9990$ \\
        $\partial^\mu S$ & $0$ & $S$ & $0.6829$ & $\Box S$ & $0.3069$ & $0.9898$ \\
        & $2$ & $\partial^\mu\partial^\nu S^{(*)}$ & $0.9814$ & & & $0.9814$ \\
        $\Box S$ & $1$ & $\partial^\mu S$ & $0.2088$ & $\Box\partial^\mu S$ & $0.7771$ & $0.9859$ \\
        $\partial^\mu\partial^\nu S^{(*)}$ & $1$ & $\partial^\mu S$ & $0.8648$ & $\Box\partial^\mu S$ & $0.1297$ & $0.9945$ \\
        \hline
        $J^\mu$ & $1$ & $\epsilon^{\mu\nu\rho}\partial_\nu J_\rho$ & $0.9747$ & & & $0.9747$ \\
        & $2$ & $\partial^\mu J^\nu$ & $0.9930$ & & & $0.9930$ \\
        $\partial^\mu J^\nu$ & $1$ & $J^\mu$ & $0.9077$ & $\Box J^\mu$ & $0.0873$ & $0.9949$ \\
        & $2$ & $\epsilon^{\mu\nu\rho}\partial^\nu\partial_\rho J_\rho$ & $0.7962$ & & & $0.7962$ \\
        $\epsilon^{\mu\nu\rho}\partial_\nu J_\rho$ & $1$ & $J^\mu$ & $0.3062$ & $\Box J^\mu$ & $0.6777$ & $0.9839$ \\
        & $2$ & $\epsilon^{\mu\nu\rho}\partial^\nu\partial_\rho J_\rho$ & $0.9207$ & & & $0.9207$ \\
        \hline
        $S'$ & $1$ & $\partial^\mu S'$ & $0.9827$ & & & $0.9827$ \\
        \hline 
        $J'^\mu$ & $0$ & $\partial_\mu J'^\mu$ & $0.9864$ & & & $0.9864$ \\
        & $1$ & $\epsilon^{\mu\nu\rho}\partial_\nu J'_\rho$ & $0.8941$ & & & $0.8941$ \\
        & $2$ & $\partial^\mu J'^\nu{}^{(*)}$ & $0.8670$ & & & $0.8670$ \\
        \hline\hline
    \end{tabular}
\end{table}

Another consistency check is the conservation of currents. Numerically, this manifests as the vanishing of the $l=0$ component of $\Lambda^z|J^\mu\rangle$ and the $l=1$ component of $\Lambda^z|T^{\mu\nu}\rangle$, which is indeed seen numerically:
\begin{equation}
    \frac{\|P_{l=0}\Lambda^z|J^\mu\rangle\|^2}{\|\Lambda^z|J^\mu\rangle\|^2}=7.2\times 10^{-4},\qquad\frac{\|P_{l=1}\Lambda^z|T^{\mu\nu}\rangle\|^2}{\|\Lambda^z|T^{\mu\nu}\rangle\|^2}=1.79\times 10^{-3}.
    \label{eq:conserv}
\end{equation}

We observe that $\partial^\mu\partial^\nu S$ and $T^{\mu\nu}$ are subject to level mixing. Due to their close scaling dimension, the second order perturbation theory strongly mix them together. To identify them, we act the conformal generator on the $\partial S$ state 
\begin{align}
    |\partial^\mu\partial^\nu S\rangle&\propto\left(|1\rangle\langle 1|+|2\rangle\langle 2|\right)\Lambda^z|\partial S\rangle\nonumber\\
    |T^{\mu\nu}\rangle&\propto\left(|1\rangle\langle 2|-|2\rangle\langle1|\right)\Lambda^z|\partial S\rangle
\end{align}
where $|1\rangle$ and $|2\rangle$ are the two lowest $l=2,s=0$ states. In Eq.~\eqref{eq:conserv}, we remove the level mixings for $T^{\mu\nu}$ in this way, but the conservation of $T^{\mu\nu}$ itself is not used as a criteria. Similar level mixing also happens to the state $\partial^\mu J'^\nu$.

\end{document}